\documentclass{PoS}

\def\reffi#1{\mbox{Figure~\ref{#1}}}

\def\citere#1{\mbox{Ref.~\cite{#1}}}
\def\citeres#1{\mbox{Refs.~\cite{#1}}}

\newcommand{\GeV}{\unskip\,\mathrm{GeV}}

\def\mathswitchr#1{\relax\ifmmode{\mathrm{#1}}\else$\mathrm{#1}$\fi}

\newcommand{\PW}{\mathswitchr W}

\newcommand{\PZ}{\mathswitchr Z}

\newcommand{\Pg}{\mathswitchr g}

\newcommand{\PH}{\mathswitchr H}
\newcommand{\Pe}{\mathswitchr e}

\newcommand{\Pd}{\mathswitchr d}

\newcommand{\Pu}{\mathswitchr u}

\newcommand{\Pp}{\mathswitchr p}
\newcommand{\Pt}{\mathswitchr t}

\newcommand{\Pqbar}{\bar{\mathswitchr q}}

\newcommand{\Pq}{\mathswitchr q}
\newcommand{\bPp}{\bar{\mathswitchr p}}

\def\mathswitch#1{\relax\ifmmode#1\else$#1$\fi}

\newcommand{\MW}{\mathswitch {M_\PW}}

\title{NLO QCD corrections to WW+jet production including leptonic W decays at hadron colliders}

\ShortTitle{NLO QCD corrections to WW+jet production}

\author{Stefan Dittmaier$^{a}$, \speaker{Stefan Kallweit}$^{b}$ and Peter Uwer$^c$
\thanks{
This work is supported in part 
by the European Community's Marie-Curie Research Training Network HEPTOOLS under contract MRTN-CT-2006-035505,
by the DFG Sonderforschungsbereich/Transregio 9 ``Computergest\"utzte Theoretische Teilchenphysik'' SFB/TR9, 
and by the Initiative and Networking Fund of the Helmholtz Association, contract HA-101 ("Physics at the Terascale").}\\
\llap{$^a$} Albert-Ludwigs-Universit\"at Freiburg, Physikalisches Institut, 
D-79104 Freiburg, Germany\\
\llap{$^b$} Paul Scherrer Institut, W\"urenlingen und Villigen,
  CH-5232 Villigen PSI, Switzerland\\
\llap{$^c$} Institut f\"ur Physik, Humboldt-Universit\"at zu Berlin,
D-10099 Berlin, Germany
\\
E-mail: \email{stefan.dittmaier@physik.uni-freiburg.de}, \email{stefan.kallweit@psi.ch}, \email{Peter.Uwer@physik.hu-berlin.de}
}
\abstract{We report on the calculation of the next-to-leading order QCD corrections
to the production of W-boson pairs in association with a hard
jet at the Tevatron and the LHC, which is an important source 
of background for Higgs and new-physics searches. 
Leptonic decays of the $\PW$ bosons are included by applying an improved 
version of the narrow-width approximation that treats the $\PW$ bosons 
as on-shell particles, but keeps the information on the $\PW$~spin.
A selection of differential NLO QCD cross sections 
is provided both for the LHC and the Tevatron. 
The QCD corrections stabilize the LO prediction for 
the cross section with respect to scale variations. 
The differential LO cross sections are generally not simply rescaled by the corrections.
Their shapes are particularly distorted if an additional energy scale is involved.
}

\FullConference{RADCOR 2009 - 9th International Symposium on Radiative Corrections (Applications of Quantum Field Theory to
Phenomenology) \\
                 October 25-30 2009\\
                 Ascona, Switzerland}

\begin{document}

\section{Introduction}
The search for new-physics particles---including the Standard Model
Higgs boson---will be the primary task in high-energy physics in the 
era of the LHC. The extremely complicated
hadron collider environment does not only require
sufficiently precise predictions for new-physics signals, but also
for many complicated background reactions that cannot entirely be
measured from data. Among such background processes, several involve
three, four, or even more particles in the final state, rendering
the necessary next-to-leading-order (NLO) calculations in QCD very
complicated. This problem lead to the creation of an
``experimenters' wishlist for NLO calculations''
\cite{Buttar:2006zd,Campbell:2006wx,Bern:2008ef} that 
were still missing at that time, but are required
for successful LHC analyses. The process
$\Pp\Pp\to\PW^+\PW^-{+} \mathrm{jet+X}$ made it to the top of this list.
Meanwhile the $2\to3$ particle processes and also some of the $2\to4$ 
particle processes~\cite{Bredenstein:2008zb,Ellis:2008qc} on the list have been
evaluated at NLO QCD. Moreover, benchmark results for the virtual corrections
have been presented for a specific phase-space point for all $2\to4$ processes
on the list in \citere{vanHameren:2009dr}.

The process of $\PW\PW$+jet production
is an important source for background to the
production of a Higgs boson that subsequently decays into a W-boson
pair, where additional jet activity might arise from the 
production. $\PW\PW$+jet production
delivers also potential background to new-physics searches, such as
supersymmetric particles, because of leptons and missing transverse
momentum from the W~decays. Besides the process is 
interesting in its own right, since W-pair production processes
enable a direct analysis of the non-abelian
gauge-boson self-interactions, and a large fraction of W~pairs
will show up with additional jet activity at the LHC.
Last but not least $\PW\PW$+jet at NLO
also delivers the real--virtual contributions 
to the next-to-next-to-leading-order (NNLO) calculation of $\PW$-pair 
production, 
for which further building blocks are presented in \citere{Chachamis:2008yb}.

Here we report on the calculation of the process 
$\Pp\Pp/\Pp\bPp\to\PW^+\PW^-{+}\mathrm{jet+X}$ in NLO QCD including 
leptonic $\PW$-boson decays. 
Results of this calculation have been published in \citeres{Dittmaier:2007th,Dittmaier:2009un}.
Parallel to our work, another NLO study~\cite{Campbell:2007ev} of 
$\Pp\Pp\to\PW^+\PW^-{+} \mathrm{jet+X}$ at the LHC appeared.
\mbox{Moreover,} NLO QCD corrections to the related processes
\mbox{$\Pp\Pp\to\PW\gamma{+}\mathrm{jet+X}$}~\cite{Campanario:2009um} and \linebreak
\mbox{$\Pp\Pp\to\PZ\PZ{+}\mathrm{jet+X}$}~\cite{Binoth:2009wk} have been
calculated recently.

\section{Details of the NLO calculation}
At leading order (LO), hadronic $\PW\PW{+}$jet production receives 
contributions from the partonic processes 
$\Pq\Pqbar\to\PW^+\PW^- \Pg$, $\Pq\Pg\to\PW^+\PW^- \Pq$, and 
$\Pg\Pqbar\to\PW^+\PW^- \Pqbar$, where $\Pq$ stands for up- or down-type 
quarks. Note that the amplitudes for $\Pq=\Pu,\Pd$ are not the same, 
even for vanishing light-quark masses. All three channels are related 
by crossing symmetry. 

\begin{figure}
\centerline{
\includegraphics[bb = 200 400 450 700, scale = .53]{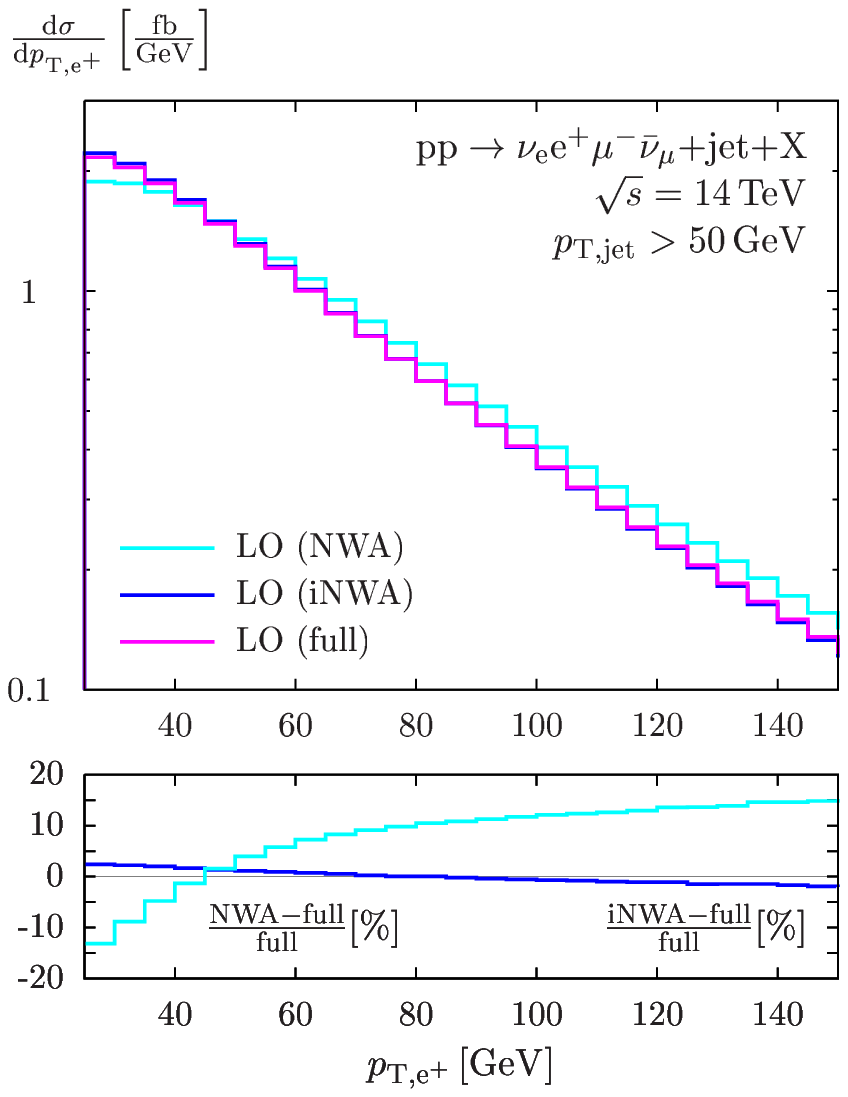}
\includegraphics[bb = 180 400 450 700, scale = .53]{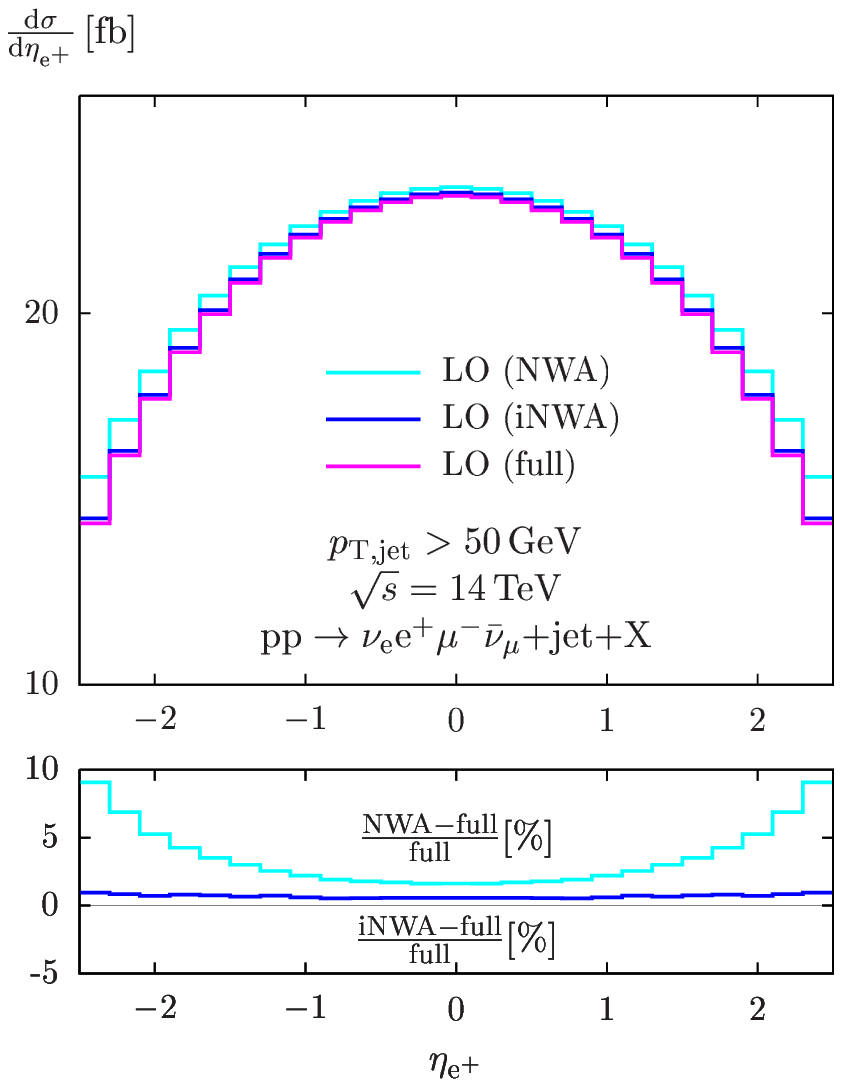}
\includegraphics[bb = 180 400 410 700, scale = .53]{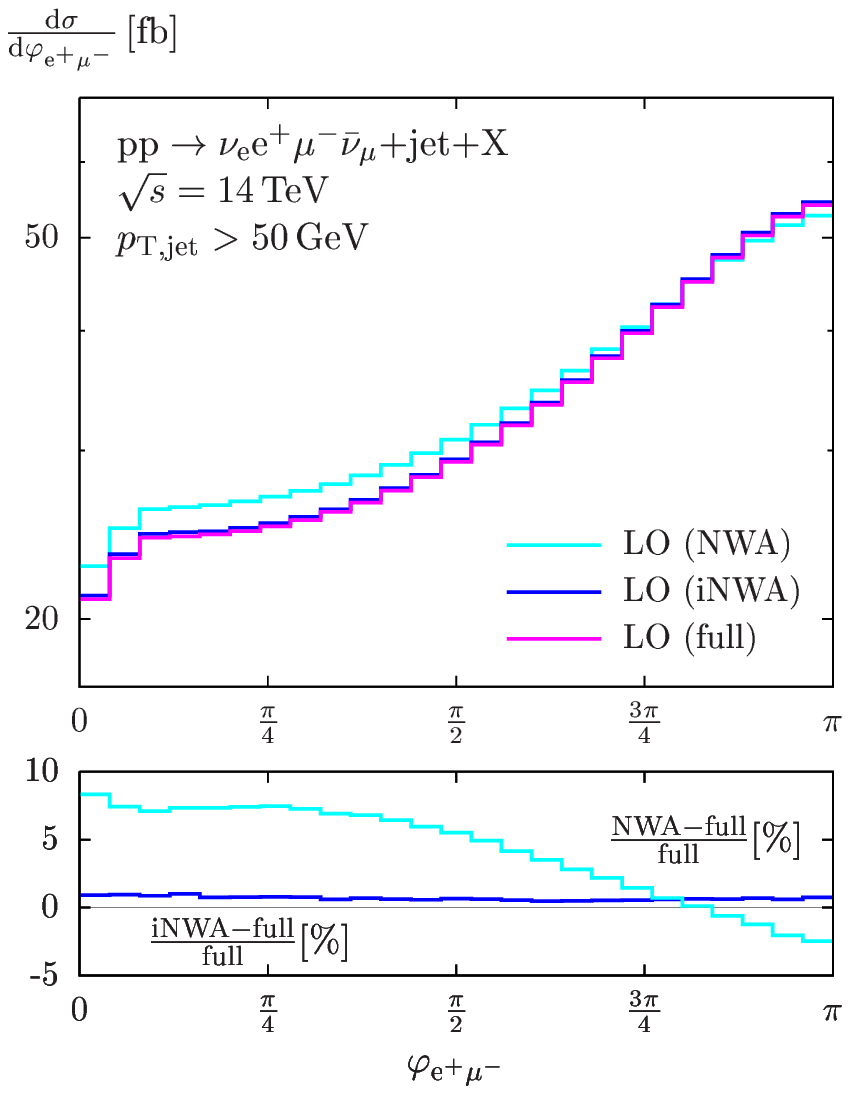}}
\caption{Comparison of $\PW$-decay descriptions in the distributions of the 
transverse momentum of $\Pe^+$ (left plot), the pseudo-rapidity of $\Pe^+$ 
(central plot), and the azimuthal angle between the two decay leptons (right plot). 
The LO cross sections are evaluated at 
\mbox{$\mu=\mu_{\mathrm{fact}}=\mu_{\mathrm{ren}}=\MW$} for the full calculation, 
the naive NWA, and the improved NWA. (Taken from \citere{Dittmaier:2009un}.)}
\label{fig:decaycomparison}
\end{figure}

The leptonic $\PW$ decays are implemented by means of an improved 
narrow-width approximation (NWA) that treats the $\PW$ bosons as on-shell 
particles, but keeps the spin correlations between production and decay processes. 
In this way, a significantly better approximation of the full calculation is 
achieved, which can be read off the comparison of the sample LO distributions 
in \reffi{fig:decaycomparison}. 

In order to prove the correctness of our results we have evaluated 
each ingredient twice using independent calculations based---as
far as possible---on different methods, 
yielding results in mutual agreement.

\subsection{Virtual corrections}
\textit{Version 1} of the virtual corrections is essentially obtained as for the related processes of $\Pt\bar\Pt\PH$
\cite{Beenakker:2002nc} and $\Pt\bar\Pt{+}$jet \cite{Dittmaier:2007wz}
production.
The Feynman diagrams are generated with 
{\sl Feyn\-Arts}~1.0 \cite{Kublbeck:1990xc}
and further processed with in-house {\sl Mathematica} routines,
which automatically create an output in {\sl Fortran}. 
The IR divergences (soft and collinear) are analytically separated
from the finite remainder in terms of triangle subdiagrams,
as described in \citeres{Beenakker:2002nc,Dittmaier:2003bc}.
This separation, in particular, allows for a transparent evaluation
of so-called rational terms that originate from $D$-dependent terms
multiplying IR divergences, which appear as single or double poles in 
$\epsilon$.
As generally shown in \citere{Bredenstein:2008zb}, after properly
separating IR from UV divergences such rational terms
originating from IR
divergences completely cancel; this general result is confirmed in our
explicit calculation.
For the results presented in \citere{Dittmaier:2007th},
the pentagon tensor integrals were directly reduced to box
integrals following \citere{Denner:2002ii}, while box and lower-point 
integrals were reduced \`a la Passarino--Veltman \cite{Passarino:1978jh} 
to scalar integrals. This procedure completely avoids inverse Gram 
determinants of external momenta in the reduction step from 5-point to 
4-point integrals, but the reduction of box and lower-point tensor
integrals involves such inverse determinants via the 
Passarino--Veltman algorithm. Although these inverse determinants
jeopardize the numerical stability in regions where such determinants
are small, sufficient numerical stability was already achieved.
Meanwhile the tensor reduction has been further improved using
the methods of \citere{Denner:2005nn}. 
The scalar one-loop integrals
are either calculated analytically or using the results of
\citeres{'tHooft:1978xw,Beenakker:1988jr,Denner:1991qq}.

\textit{Version 2} of the evaluation of loop diagrams starts
with the generation of diagrams and amplitudes via 
{\sl Feyn\-Arts}~3.4 \cite{Hahn:2000kx}
which are then further manipulated with {\sl FormCalc}~6.0
\cite{Hahn:1998yk} and eventually
automatically translated into {\sl Fortran} code.
The whole reduction of tensor to scalar integrals is done with the
help of the {\sl LoopTools} library \cite{Hahn:1998yk},
which employs the method of \citere{Denner:2002ii} for the
5-point tensor integrals, Passarino--Veltman \cite{Passarino:1978jh}
reduction for the lower-point tensors, and the {\sl FF} package 
\cite{vanOldenborgh:1989wn,vanOldenborgh:1991yc} for the evaluation 
of regular scalar integrals.
The dimensionally regularized soft or collinear singular 3- and 4-point
integrals had to be added to this library. To this end, the
explicit results of \citere{Dittmaier:2003bc} for the vertex and of 
\citere{Bern:1993kr}
for the box integrals (with appropriate analytical continuations)
are taken.

\subsection{Real corrections}
The matrix elements for the real corrections are given by the processes
$0 \to \PW^+\PW^-   \Pq \bar \Pq \Pg \Pg$ and
$0 \to \PW^+\PW^-   \Pq \bar \Pq \Pq' \bar \Pq'$
with a large variety of flavour insertions for the light quarks
$\Pq$ and $\Pq'$.
The partonic processes are obtained from these matrix elements 
by all possible crossings of quarks and gluons into the initial state. 
The evaluation of the real-emission amplitudes is
performed in two independent ways. In one approach we apply the 
Weyl--van-der-Waerden formalism (as described in
\citere{Dittmaier:1998nn}). The other one is based on {\sl Madgraph} 
\cite{Stelzer:1994ta} generated code.
Both evaluations employ 
(independent implementations of) the dipole subtraction formalism 
\cite{Catani:1996vz}
for the extraction of IR singularities and for their
combination with the virtual corrections. 

In one calculation the phase-space integration is performed by a multi-channel 
Monte Carlo integrator~\cite{Berends:1994pv} with weight 
optimization~\cite{Kleiss:1994qy} written in {\sl C++}, which 
is constructed similar to {\sl RacoonWW}
\cite{Denner:1999gp,Roth:1999kk}. The second calculation uses a 
simple mapping where the phase space is generated from 
a sequential splitting.

\section{Numerical results}
We consistently use the CTEQ6 \cite{Pumplin:2002vw}
set of parton distribution functions (PDFs), i.e.\ we take
CTEQ6L1 PDFs with a 1-loop running $\alpha_{\mathrm{s}}$ in
LO and CTEQ6M PDFs with a 2-loop running $\alpha_{\mathrm{s}}$
in NLO. The complete setup we used for our numerical calculations is precisely defined in 
\citere{Dittmaier:2009un}, where a large variety of additional results is provided.

\begin{figure}
\centering
\includegraphics[bb = 200 400 450 650, scale = .53]{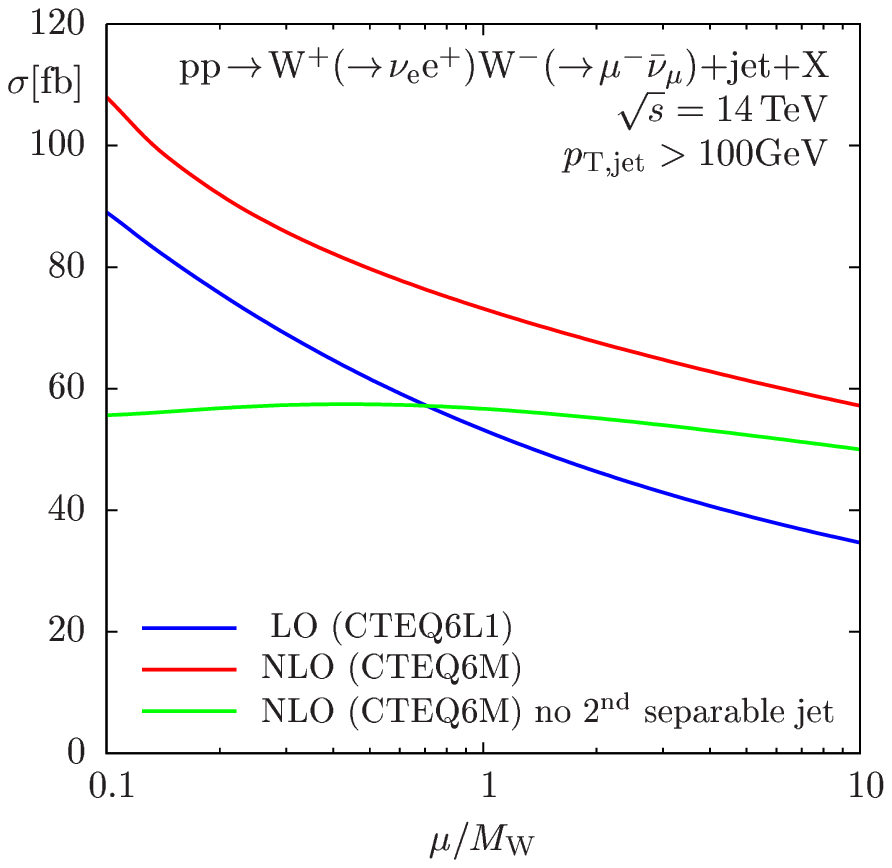}
\includegraphics[bb = 160 400 410 650, scale = .53]{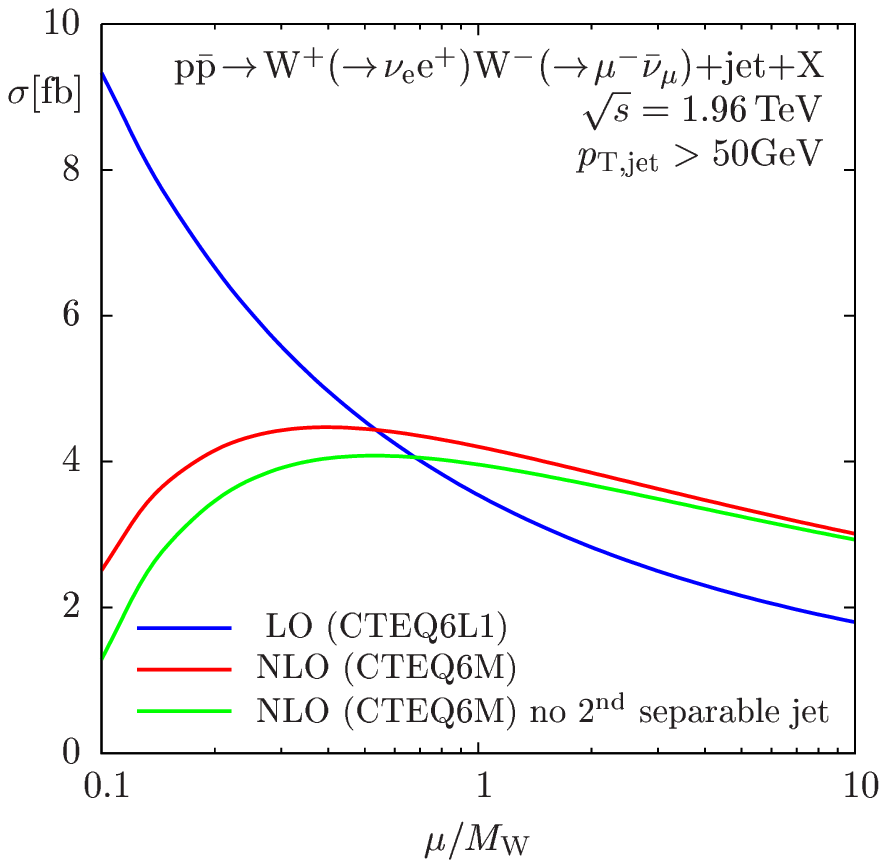}
\caption{Scale dependence of the WW+jet cross sections with W decays included 
and further cuts applied according to \citere{Dittmaier:2009un}. 
For the LHC setup, the results are given for $p_{\mathrm{T,jet}}>100\GeV$ 
(left plot). For the Tevatron we show results for  $p_{\mathrm{T,jet}}>50\GeV$ 
(right plot).}
\label{fig:scalevariationdecay}
\end{figure} 

\begin{figure}
\centerline{
\includegraphics[bb = 200 400 450 700, scale = .53]{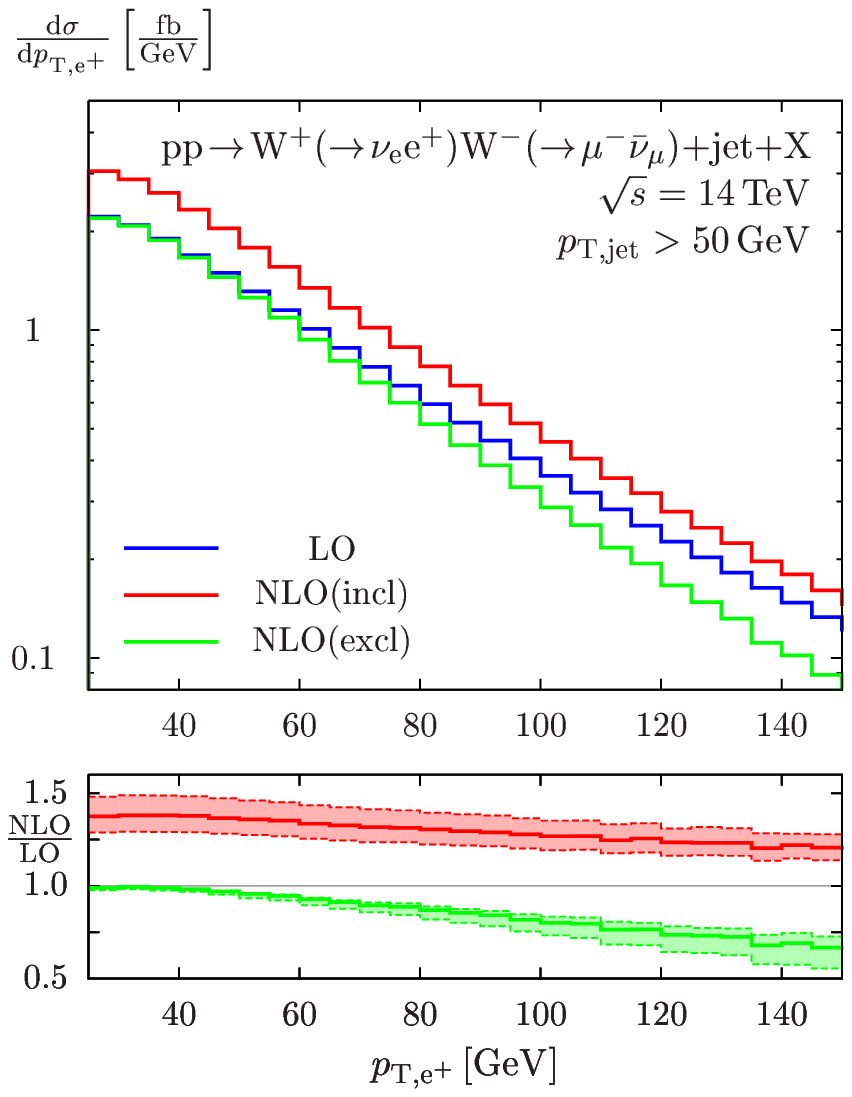}
\includegraphics[bb = 180 400 450 700, scale = .53]{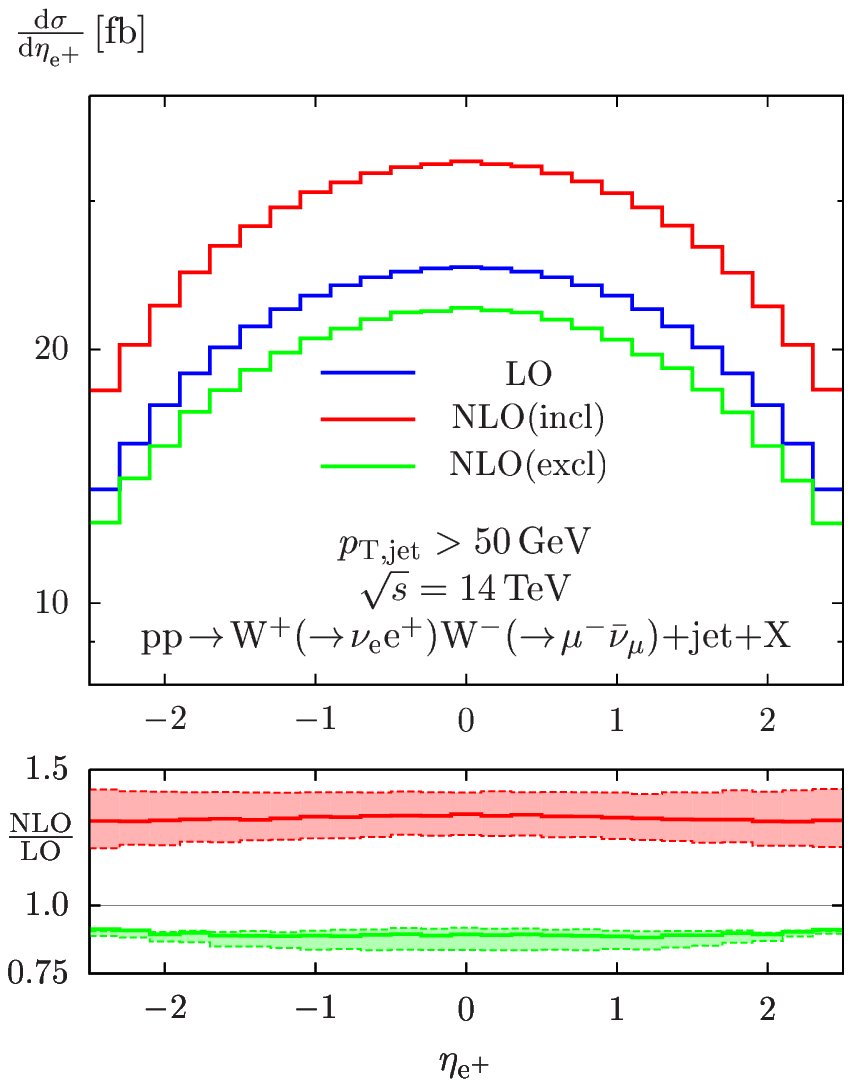}
\includegraphics[bb = 180 400 410 700, scale = .53]{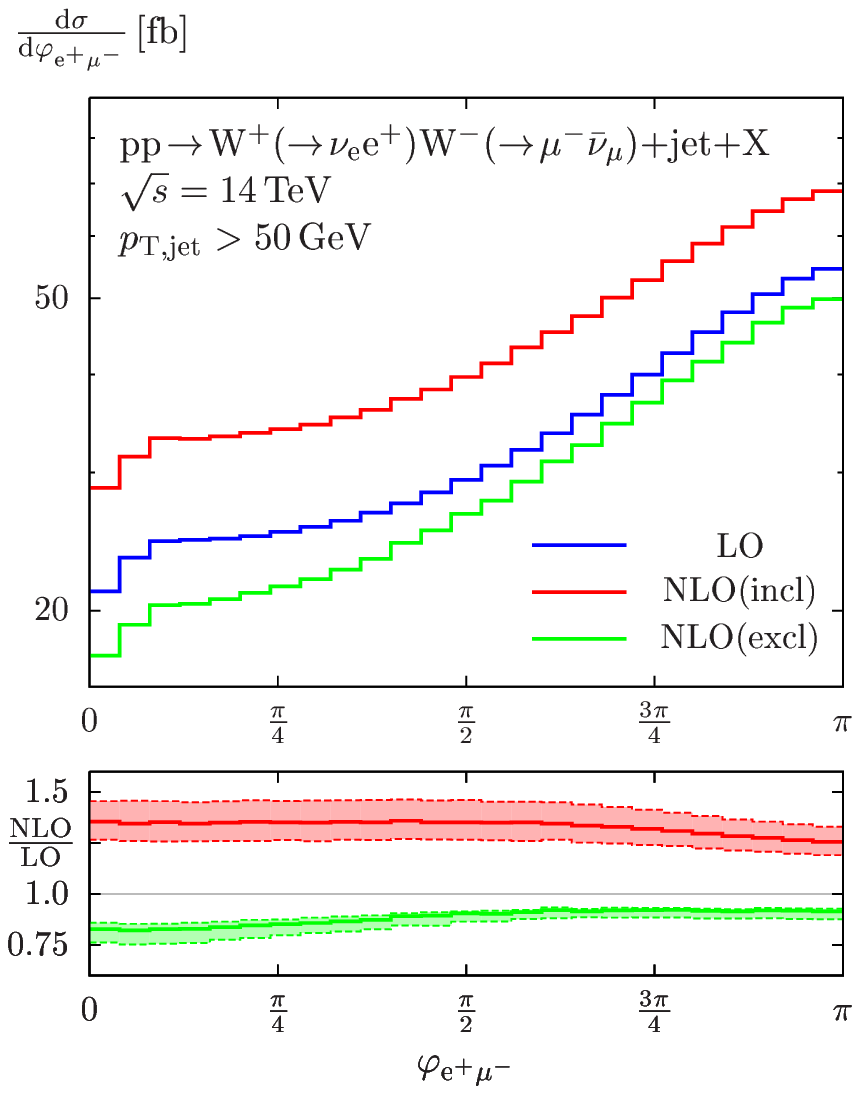}}
\caption{Differential cross sections for WW+jet with decays included in the
  improved NWA at the LHC: The LO and NLO distributions are shown for
  \mbox{$\mu=\mu_{\rm{fact}}=\mu_{\rm{ren}}=\MW$}. The distributions of the transverse 
momentum of $\Pe^+$ (left plot), the pseudo-rapidity of $\Pe^+$ (central plot), and the azimuthal angle between the 
two decay leptons (right plot) are depicted. The bands in the $K$-factors refer to a variation 
of $\mu$ by a factor of 2 in the NLO quantities.
(Taken from \citere{Dittmaier:2009un}.)}
\label{fig:distributions}
\end{figure}

\reffi{fig:scalevariationdecay} shows the scale dependence of the NLO cross section
for the LHC and the Tevatron.
The QCD corrections stabilize the LO prediction for
the WW+jet cross section considerably with respect to a
variation of the factorization and renormalization scales which we
identify with each other. At the LHC, this stabilization
of the prediction, however, requires a veto on a second hard jet.
Otherwise the production of final states with WW+2jets, which 
yields a LO component of the NLO
correction, introduces again a large scale dependence.
In \reffi{fig:distributions} a sample of NLO distributions is provided for the LHC setup.
At the LHC the psuedo-rapidity distributions in the dominant region and also the
distributions in the angles between the two charged leptons
have an almost constant
$K$-factor of about 1.3 (inclusive cross-section definition);
for the exclusive cross-section definition the corrections are even smaller
and rather close to 1. The $p_{\mathrm{T}}$ spectra, on
the other hand, show a much more phase-space-dependent $K$-factor
with the exclusive cross-section definition showing an even larger
dependence than the inclusive one. This is not surprising since the
$p_{\mathrm{T}}$ introduces an additional scale which could introduce
potentially large logarithms which are badly treated by a constant
renormalization scale.
At the Tevatron our findings are similar~\cite{Dittmaier:2007th,Dittmaier:2009un}.
We note that the almost constant $K$-factor which holds for a remarkable number 
of distributions has also been observed in \citere{Campbell:2007ev}.

\end{document}